\documentclass[twocolumn]{aa} 

\usepackage[varg]{txfonts}
\usepackage{graphicx}
\usepackage{CJK}

\authorrunning{Li et al}
\titlerunning{New detection of blue straggler stars in 50 open clusters}

\begin{document} 
\begin{CJK}{UTF8}{gbsn}

    \title{The new detection of blue straggler stars in 50 open clusters using Gaia DR3}

    \author{Chunyan Li (李春燕）\inst{1,2}
          \and
          Jing Zhong （钟靖）\inst{1}
          \and
          Songmei Qin (秦松梅) \inst{1,2}
          \and
          Li Chen （陈力）\inst{1,2}}

\institute{ Key Laboratory for Research in Galaxies and Cosmology, Shanghai Astronomical Observatory, Chinese Academy of Sciences, 80 Nandan Road, Shanghai 200030, China \\
             \and
              School of Astronomy and Space Science, University of Chinese Academy of Sciences, No. 19A, Yuquan Road, Beijing 100049, China. \\
             {\it e-mail: lichunyan@shao.ac.cn,chenli@shao.ac.cn} }
   \date{Received 16, September 2022; accepted 13, January 2023}

\abstract
{The particularly abundant presence of blue straggler stars (BSS) in Galactic open clusters offers favorable conditions for detailed studies on the statistical properties and the origin of the blue straggler population. With the help of Gaia DR3, the number of identified open clusters continuously increases, and the determination of star cluster members is more reliable.}
{We performed a more thorough search for BSS in newly found open clusters based on Gaia data.}
{We implemented a uniform membership determination for over one thousand newly identified open clusters with larger sky coverage based on the astrometric and photometric data from Gaia DR3. The membership probabilities of stars were assigned by the pyUPMASK algorithm. Then we estimated the physical parameters of these clusters by isochrone fitting on their CMDs and picked out BSS in the specific region of these CMDs.}
{We identified 138 BSS that had not been reported before in 50 open clusters.}
{Compared with recent catalogs that present more than 1500 BSS in 339 open clusters, our new catalog increased the number of BSS in Galactic open clusters by about 10\%, and the number of open clusters with BSS by nearly 17\%. In the future, more accurate abundance measurements are anticipated to better probe the origin of BSS in open clusters.}

 
\maketitle
\section{Introduction}
Blue straggler stars (BSS) were initially identified by \cite{Sandage_1953} in the color--magnitude diagram (CMD) of the globular cluster M3, appearing as an extension of the cluster main sequence, blueward and above the main sequence turnoff (MSTO). 
BSS are found in globular clusters in the Milky Way \citep{Piotto_2004,2008ApJ...678..564L} or dwarf spheroidal galaxies \citep{2012MNRAS.421..960S}, and also commonly in the intermediate-age to old open clusters \citep[hereafter Rain21, Jadhav21]{Rain_2021,Jadahav2021} of our Galaxy.
An open cluster (OC) is a group of stars that are born from the collapse of the same dense molecular cloud, and therefore members in an OC share a common distance and kinematics as well as initial chemical composition and age, thus providing a good test bed for studying the theories of stellar evolution.
The abundant presence of BSS in Galactic OCs poses a challenge for standard simple-star population (SSP) theory \citep{2005ApJ...619..824X, 2007ApJ...660..319X}.  

There are multiple formation scenarios for creating BSS in OCs. 
Mass transfer from binary companions can lead to a rejuvenation of the acceptors and the formation of BSS \citep{1964MNRAS.128..147M}. Individual stellar mergers resulting from direct stellar collisions are also linked to BSS formation \citep{1976ApL....17...87H, 1989AJ.....98..217L}. 
\cite{Andronov_2006} quantified the angular momentum loss induced by magnetic stellar winds in main sequence tidally synchronized binaries and suggested that this mechanism is responsible for at least one-third of the BSS in OCs older than 1 Gyr. It was also proposed in recent studies that BSS formed from the merger of main sequence stars previously in a hierarchical triple system as a result of the eccentric Kozai--Lidov mechanism \citep{Perets_Fabrycky_2009, Naoz_2014}, which has a significant role in BSS formation in OCs.

For detailed studies on the statistical properties and the origin of blue straggler population, we require a comprehensive and homogeneous catalog of blue stragglers in OCs.
As most Galactic OCs are located on the thin disk, observations of OCs are often hampered by severe contamination from dominant background or foreground field stars, which leads to more uncertainties in the characterization of cluster properties.
Before the Gaia era, it was rather difficult for reliable member star selection due to limited astrometric precision, bringing about inconsistency in the determination of basic parameters, such as the distance, kinematics, and age of OCs \citep{2015A&A...582A..19N}.
Based on the WEBDA database, \citet[hereafter AL07]{Ahumada_2007} identified 1887 blue straggler candidates in 199 OCs, of which 200 ($10.6\%$) stars were classified as reliable blue straggler samples.
However, lacking high precision astrometric data at the time AL07 was published, the membership estimation of stars in an OC is usually based on photometric data only, and in many cases with lower reliabilities.

Open cluster members exhibit greater clustering characteristics than field stars in spatial and kinematic spaces.The higher precision proper motion and parallax from Gaia data provide an important advantage for distinguishing cluster members from field stars.
The Gaia DR2 catalog \citep{gaia18} presents more than 1.3 billion stars with unprecedented high-precision astrometric and photometric data, greatly improving the reliability of stellar membership determination and characterization of a large number of OCs.
Based on Gaia DR2 data, \cite{Cantat-Gaudin_2018} used the UPMASK algorithm to select cluster members and provided an updated catalog of 1229 OCs, including previously known clusters and 60 newly discovered clusters.
Later on, \citet[hereafter CG20a]{Cantat-Gaudin_2020} obtained more cluster samples while eliminating a number of nonphysical groupings that had been incorrectly cataloged, and finally compiled a list of members for 1481 clusters.
Rain21 updated the BSS sample in AL07 by utilizing members in the 1481 cluster listed in CG20a. To probe BSS in the cluster, all corresponding CG20a members with probabilities of $ \rm P_{memb}\geq 50\%$ were plotted in the G versus ($ \rm G_{BP}-G_{RP}$) diagram.
The blue straggler candidates were singled out in the corresponding region of the cluster CMD.
In their result, Rain21 excluded 39 cluster samples in AL07, and added 19 recently discovered objects, eventually identifying 897 BSS in a total of 111 OCs. 

Soon after the CG20a work, \citet[hereafter CG20b]{2020A&A...640A...1C} updated a comprehensive list of 1867 bona fide OCs confirmed by analysis based on Gaia data, providing a homogeneous catalog of OC properties and a catalog of cluster member stars.
In a parallel work of Rain21, Jadhav21 produced a catalog of BSS using CG20b data, identified 868 BSS candidates in 228 clusters and 500 probable BSS (pBS) in 208 clusters. 
The difference between their catalog and the Rain21 work could come from the adopted age criteria, selection methods, and different membership probability cutoffs used in the two studies.
Although claims have been made that the sample of known open clusters might be nearly complete out to a distance of 1.8 kpc \citep{2013A&A...558A..53K, 2018A&A...615A..12Y}, it is anticipated now that more discoveries of new OCs will be in the ascendant by utilizing high-precision Gaia astrometric and photometric data. 
For example, \citet{Ferreira2020} and \citet{2021MNRAS.502L..90F} reported the discovery of 59 new OC candidates in total as a result of systematic searches, mostly toward the Galactic bulge, using Gaia DR2 data. About $75\%$ of the newly discovered clusters are older than 100 Myr.
\cite{2021A&A...646A.104H}, after a detailed inspection of multiple clustering approaches, announced the finding of 41 new OCs by applying the more efficient HDBSCAN algorithm, also based on Gaia DR2 data. 

As the early stage of Gaia's third data release, Gaia EDR3 \citep{2021A&A...649A...1G} provides astrometric and photometric parameters of 1.5 billion sources with higher accuracy than Gaia DR2.
In our recent work \citep{2022ApJS..260....8H}, we carried out a blind search for the new OCs in the solar neighborhood using Gaia EDR3 data \citep{2021A&A...649A...1G}.
After dividing low galactic latitude regions into $2\times 2$ degrees ($ \rm |b|<5$ deg) or $3\times 3$ degrees ($ \rm |b|>5$ deg) blocks, we performed a DBSCAN clustering process in the stellar five-dimension (position, proper motion, and parallax) phase space to obtain an initial sample of OC candidates. 
Subsequently, the membership probabilities for stars in the candidate cluster region were calculated by pyUPMASK.
After the visual investigation of the CMDs, we singled out 541 star clusters unknown in the literature. 

Compared to Gaia DR2, Gaia DR3 provides precision that is two to three times better for proper motion and over 20\% better for parallax parameters. In the present work we searched for BSS in OCs based solely on the Gaia DR3.
To establish a more complete BSS sample of Galactic OCs, we systematically searched for BSS in all the 2975 genuine OCs solely recognized with Gaia data \citep{2019JKAS...52..145S,2019ApJS..245...32L,2020PASP..132c4502H,2020MNRAS.499.1874M,2020A&A...635A..45C,2021RAA....21...45Q,2021RAA....21...93H,2021RAA....21..117C,2021A&A...646A.104H,2021MNRAS.502L..90F,2022ApJS..260....8H}.
We applied pyUPMASK for membership assessment with five dimensions (position, proper motion, and parallax) of Gaia data, then chose members with probability $ \rm P_{memb}> 50\%$ in the CMD, and proceeded with isochrone fitting using the Padova theoretical library. By examining the location of member stars in the CMD, we identified blue straggler member stars in the OC. Consequently, we identified 138 BSS in 50 OCs which had never been reported before. 

In Sect. \ref{sec:data_method} we describe in detail the process used to search for and authenticate the BSS candidates in OCs. In Sect. \ref{sec:result} we provide the resultant basic properties of OCs with newly identified BSS. In Sects. \ref{sec:discus} and \ref{sec:summary} we briefly discuss and summarize our results.
\section{Data and method}\label{sec:data_method}

To date, the most extensive surveys of BSS candidates in OCs have been performed by Rain21 and Jadhav21, both on the basis of Gaia DR2.
Rain21 identified BSS in 111 OCs based on the cluster members provided by CG20a. At nearly the same time, Jadhav21 picked out BSS candidates in 228 OCs based on a larger cluster catalog from CG20b. In the meantime, as many as one thousand new OCs have been reported in the literature \citep{2019JKAS...52..145S,2019ApJS..245...32L,2020PASP..132c4502H,2020MNRAS.499.1874M,2020A&A...635A..45C,2021RAA....21...45Q,2021RAA....21...93H,2021RAA....21..117C,2021A&A...646A.104H,2021MNRAS.502L..90F,2022ApJS..260....8H} outside the OC list of CG20a and CG20b.
In order to obtain a more complete sample of BSS in OCs, we hunted for BSS mainly outside the list of open clusters cataloged in Rain21 and Jadhav21.
\subsection{Initial cluster sample}\label{subsec:initial_cluster}
We collected 1001 newly found OCs based on Gaia data. Among these new OC samples, which are not included in the catalogs of CG20a or CG20b, a total of 92715 member stars are present in 894 OCs.
In addition, we note that 71 OCs listed in \cite{2019JKAS...52..145S} and 36 more OCs reported by \cite{2020PASP..132c4502H} and \cite{2021RAA....21..117C} have no member star information available. These 107 clusters are also included in our initial sample. 

For the 894 clusters with member star information in the literature, we examined their CMDs (G vs. $\rm G-G_{RP}$) with membership probability $P_{memb}> 0.5$ (or all stars if no membership information was available) one by one. By visual inspection of the main sequence morphology in the CMD and combining it with the age result from the literature, we filtered out young clusters and retained 179 OCs with log(age) over 6.8 as part of the initial sample for further BSS exploration (see Sect. \ref{sec:search_bss} for details). 
\subsection{Data preprocessing}
\label{sec:preprocessing}
We downloaded the Gaia DR3 data of cluster samples according to their coordinates, proper motion, and parallax parameters listed in the literature.
We set the radius of the projected area to 20 pc for all cluster samples.
We also limited the $\rm G$mag to 19 mag for the downloaded stars. This will significantly reduce the faint background sources. On the other hand, referring to the online description of limitations of Gaia DR3\footnote{https://www.cosmos.esa.int/web/gaia/dr3} for faint sources (Gmag > 19 mag), the quality of their data can be problematic, and the errors reported for these stars may be larger in comparison with the brighter ones. The typical uncertainties of position, parallax, and proper motions at Gmag = 20 mag are 0.4 mas, 0.5 mas, and 0.5 mas/yr respectively, and the photometric uncertainties at Gmag = 20 mag are around 1 mmag, 108 mmag, and 52 mmag for the G, BP, and RP filters, respectively. 
The most efficient way of hunting out clusters is to look for the clustering of stars in the velocity space. 
To highlight the cluster members in the proper motion distribution diagram, we cut the downloaded data according to their 3D spatial coordinates, and selected different parallax ranges in different cases: 1) if d \textless\ 2 kpc, we downloaded data in the parallax range 0.4 \textless\ parallax \textless\ 1.5mas; 2) if 2 \textless\ d \textless\ 3 kpc, data in the range 0.2 \textless\ parallax \textless\ 1 mas were selected; 3) if 3 \textless\ d \textless\ 4 kpc, data in the range 0.1 \textless\ parallax \textless\ 0.7 mas were downloaded; 4) for some distant open clusters (d \textgreater\ 4 kpc), the download data have 0.1 \textless\ parallax \textless\ 0.5 mas.
Where d is the adopted cluster distance from the literature.
Before the membership determination, we also limited the input data to a circular region centered on the published average proper motion with a 5$\sigma$ radius to remove the noticeable background and foreground field stars. About 1$\%$ of the download data are retained for each cluster as the input data of the following membership estimation.
\subsection{Membership estimation}
\label{sec:Membership}
In this work we used the pyUPMASK \citep{2021A&A...650A.109P} algorithm to perform a membership estimation for each sample star.
pyUPMASK is an open-source software package compiled by the Python language following the development principle of UPMASK\footnote{Unsupervised Photometric Membership Assignment in Stellar Clusters} \citep{2014A&A...561A..57K}, which is a member star determination method first developed to process photometric data, though has also been widely used in the determination of member stars based on astrometric parameters \citep{Cantat-Gaudin_2018, Cantat-Gaudin_2020}.
The key assumption of UPMASK is that cluster members are spatially more crowded and have similar properties (such as concentrated proper motion and parallax), while field stars are more random and uniform in space. 

In order to obtain the membership probability of cluster members more effectively and quickly, pyUPMASK improves the clustering method, member determination step, and membership probability assignment method in its determination process.
In the process of determining membership probabilities, this algorithm identifies the samples as
members or nonmembers via five-dimensional parameters ($\mu_{\alpha^\ast}$, $\mu_{\delta}$, $\varpi$, RA, and DEC). There are 18 clustering methods available; we used the Mini Batch K-Means algorithm, which is a method of data clustering according to the distance from each center point. In order to reduce the inconsistency of clustering results due to the randomness of the initial center point assignment, the pyUPMASK uses the same input data to perform the multiple allocation and decision process. Finally, each star has multiple decision results, and the member probability is the frequency of stars marked as members ($P_{memb} = n / N$, where N is the repeating times, and n is the number of times when a single star is determined as a member). 
\subsection{Field star contamination}
\label{sec:contamination}
It is worth noting that the pyUPMASK algorithm performs well in distinguishing field stars. According to the evaluation results of the UPMASK algorithm \citep{2014A&A...561A..57K}, the contamination rate increases with the increase in the sample's contamination index (CI), the ratio of the total number of field stars to the number of cluster members. Specifically, in the absence of Gaia data in their work, the membership determination mainly comes from the photometric data and the average contamination rate is about 10$\%$ for a sample with a CI less than 1000. With the help of the accurate astrometric data provided by Gaia, we can now obtain better membership determination results. Moreover, \cite{2021A&A...650A.109P} pointed out that pyUPMASK has better performance than UPMASK in a variety of statistical indicators, especially for the samples including proper motion parameters. 

According to the probability function adopted by pyUPMASK, field stars are usually assigned lower member probability values. 
We investigated the properties of stars with $P_{memb} < 0.5$ and found that they present a uniform distribution in position space and have no obvious clumping structure in the proper motion space. On the other hand, we also checked the histogram of membership probability of all output data, which showed that the number of stars between probability 0.5 and 0.9 is relatively small. 
We wanted to use as many stars as possible; therefore, we selected stars whose membership probabilities are greater than 0.5 as cluster members. The criteria probability of 0.5 is also consistent with the criteria of other works (e.g., \citealt{2019A&A...626A..17C,Yontan_2019,2021Ap&SS.366...68A}). 
\subsection{Identification of BSS}\label{sec:search_bss}
Blue stragglers stand out from other cluster members due to their position in the CMD that is bluer and brighter than the MSTO region \citep{Sandage_1953}.
Following this definition, we searched for blue stragglers in a specific region of the cluster CMD, which was determined by the procedure described below. 

Differential reddening (DR) correction. Considering that the DR introduces some dispersion at the CMD position, especially near MSTO, which can affect the selection of BSS, we first referred to the method of \cite{Milone_2012} to correct the DR of the member stars for each open cluster. Then the corrected G vs. ($\rm G-G_{RP}$) diagram was drawn. 

Isochrone fitting. We then used the theoretical isochrone from PARSEC \citep{Bressan_2012} to perform the CMD fitting. The chosen range of age and metallicity for isochrones is $ \rm logt \in (6.0, 10.0) $ and $ \rm Z \in (0.005, 0.03) $, with step sizes of 0.05 and 0.002, respectively. We visually inspected the match of the isochrones to the significant characteristic regions, such as the main sequence, the turn-off, and the red giant or red clump features in the CMDs, and derived the age, reddening E(G-RP), distance module (DM), and metallicity parameters for each open cluster from the best fitting isochrone. 

\begin{figure}[!htbp]
    \centering
    \includegraphics[width=\linewidth]{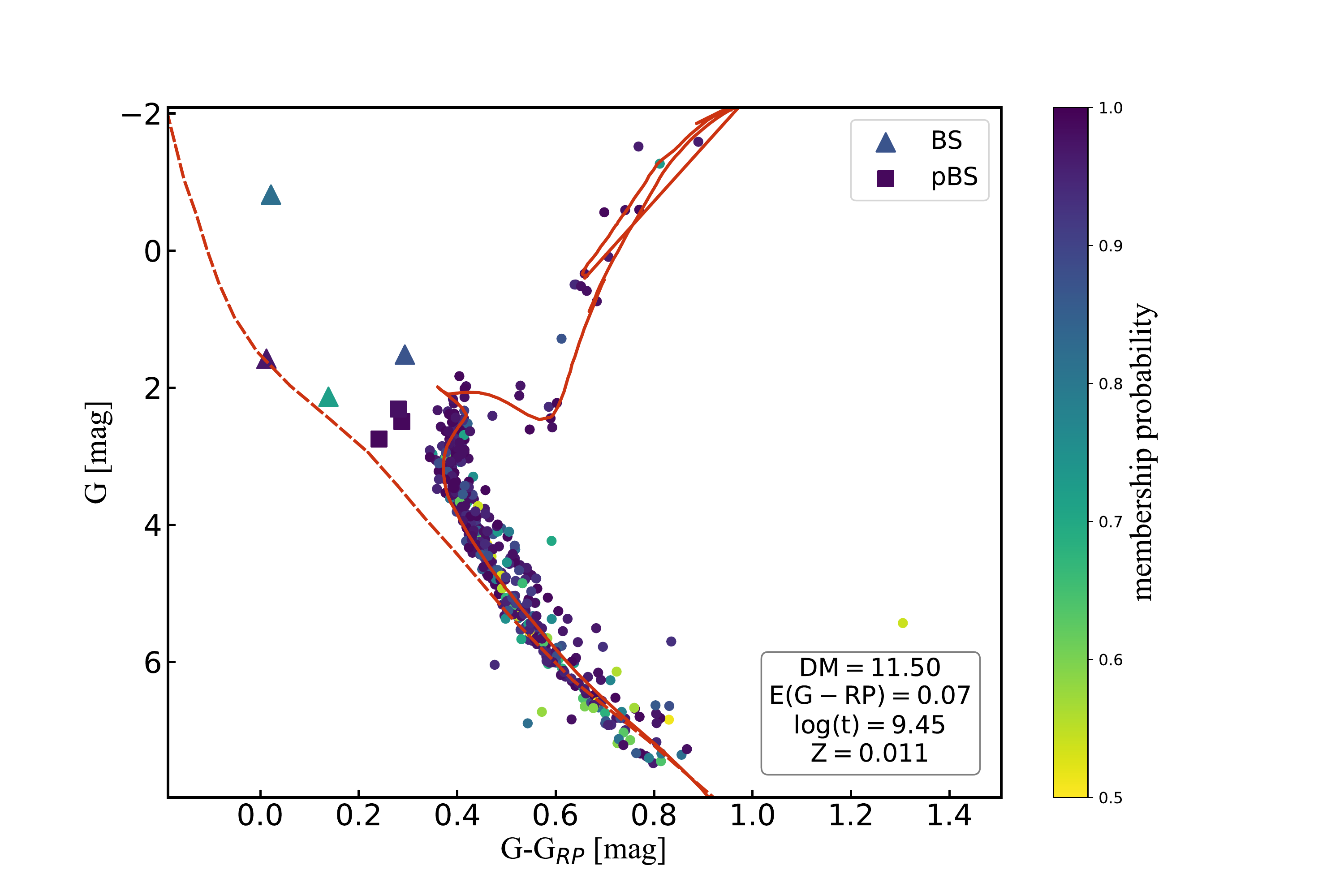}
    \caption{Differential reddening corrected CMD of UBC 274 with isochrone and ZAMS. Dots with different colors are cluster members with different membership probabilities, but all have $ \rm P_{memb}\geq 0.5$. Triangles and squares are respectively the BS and pBS identified in this work. Red solid and dashed lines are the best fitting isochrone and ZAMS, respectively. The derived parameters of UBC 274 from the best fitting isochron are $ \rm log(t)=9.45$, $ \rm Z=0.011$, $ \rm DM=11.5$, $ \rm E(G-RP)=0.07$.}
    \label{fig:CDM_class1}
\end{figure}

\begin{figure}[!htbp]
    \centering
    \includegraphics[width=\linewidth]{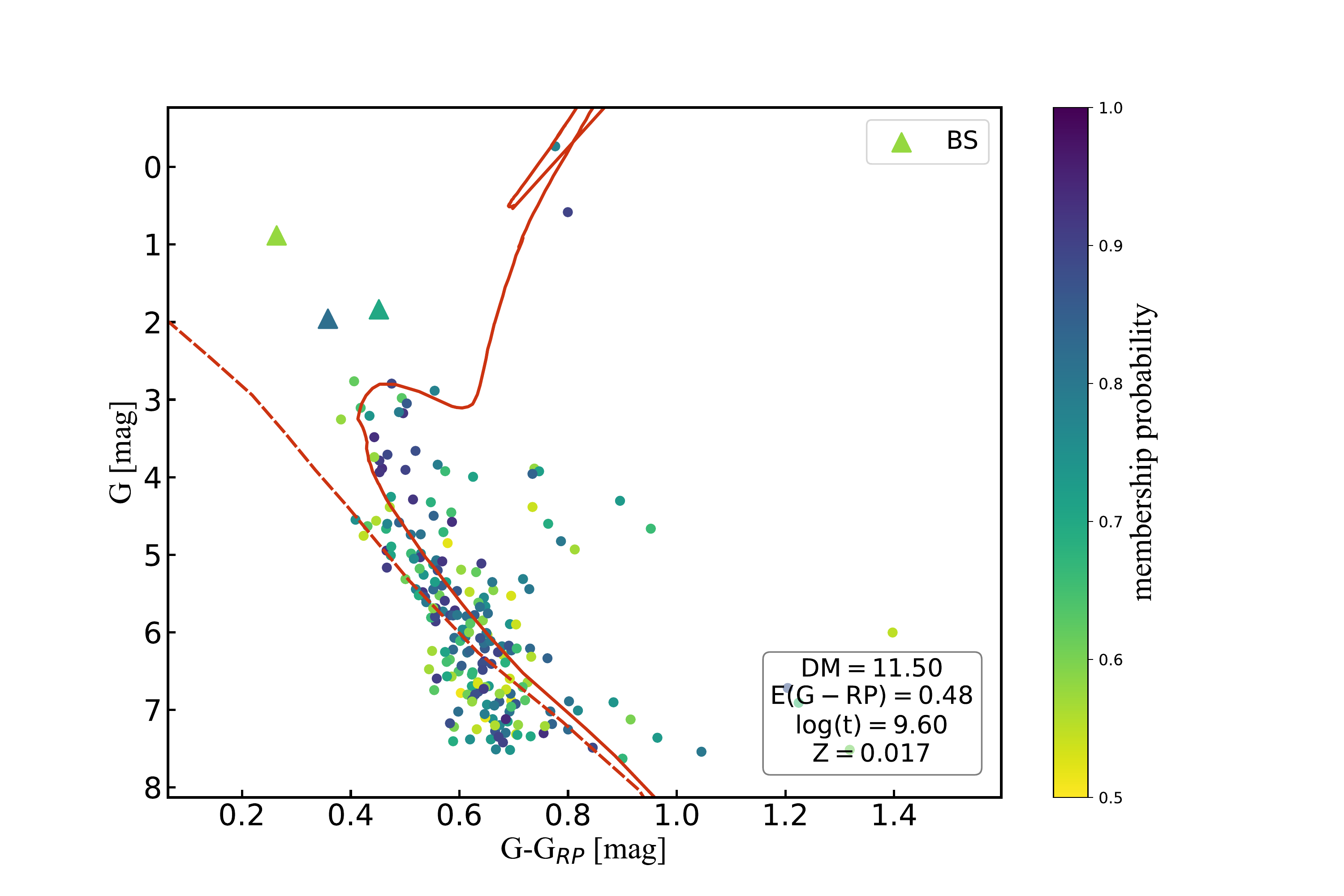}
    \caption{Same as Fig. \ref{fig:CDM_class1} but for cluster CWNU 70.}
    \label{fig:CDM_class2}
\end{figure}

BSS identification. Referring to Rain21, the genuine or possible blue stragglers can be in the region roughly delimited by the isochrone and zero age main sequence (ZAMS) as red solid and dashed lines in Fig. \ref{fig:CDM_class1}, respectively. In view of the different reliabilities, the stars in the region were divided into two classes. The stars closer to the main sequence turn off point were classified as possible blue straggler stars (pBS; squares in Fig. \ref{fig:CDM_class1}), and the remaining stars (triangles in Fig. \ref{fig:CDM_class1} and \ref{fig:CDM_class2}) were identified as genuine blue straggler stars (BS).

\subsection{Yellow stragglers}\label{sec:yss}
Yellow straggler stars (YSS) have colors between those of the TO and the RGB, and brighter than the subgiant branch \citep{2004AJ....128.3019C}.
A YSS could be the pairing of a post main sequence star (which is more massive than a TO star and on its way to the RGB) and an evolved BSS \citep{1990AJ....100.1859M}. However, the YSS region can also contain the binaries’ product of mass transfer or mergers \citep{2006A&A...455..247T,2008MNRAS.384.1263C}.
The methodology followed to find YSS is the same as that in \cite{Rain_2021} (their Fig. 3).
A total of 14 YSS candidates were recognized in 7 of our 50 sample clusters.
\section{Results}
\label{sec:result}
\begin{figure}[!htbp]
    \centering
    \includegraphics[width=\linewidth]{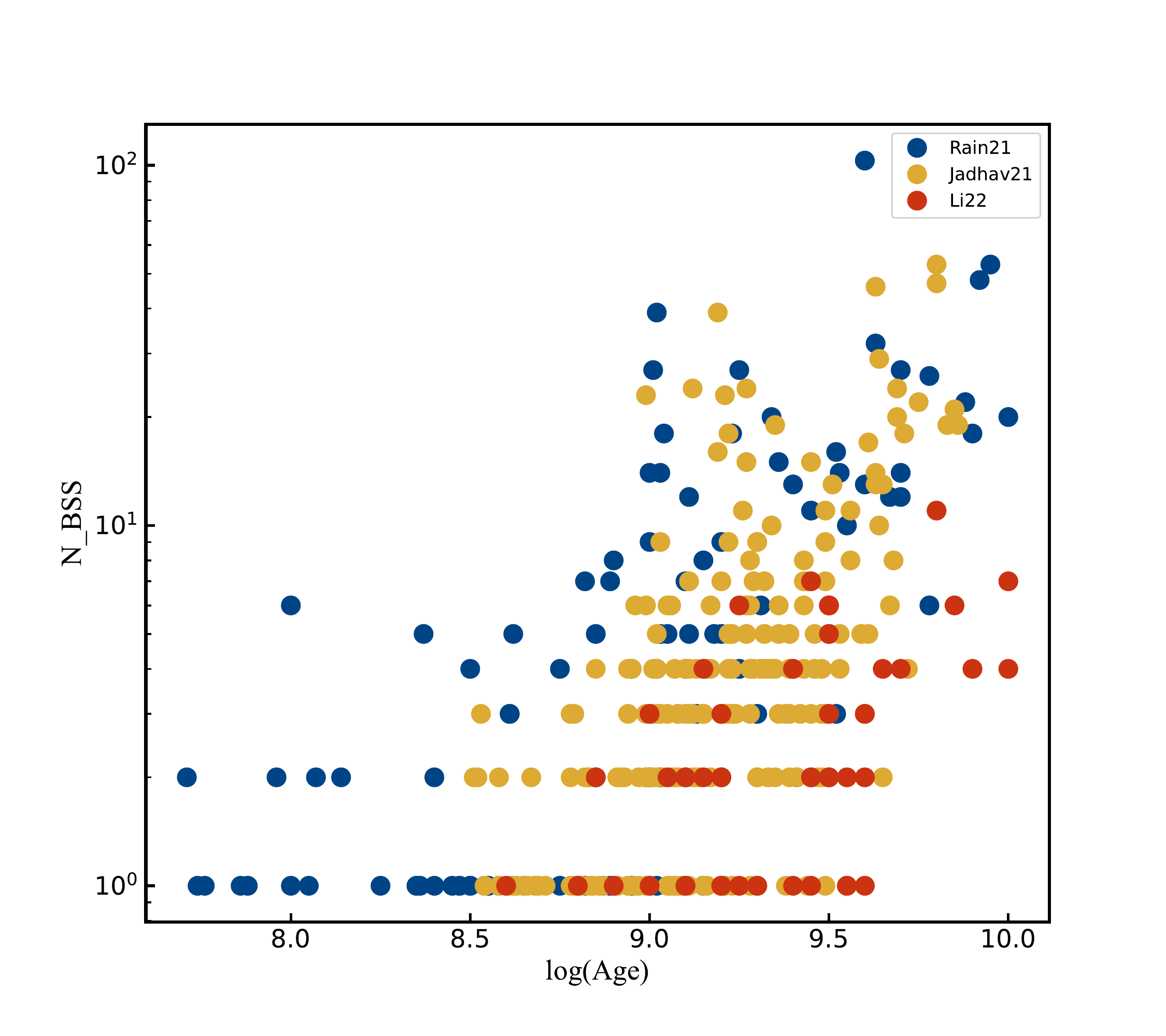}
    \caption{Relationship of the number of blue stragglers in different works on the age of clusters. Red dots represent the BS+pBS sample in this work, yellow dots represent the BSS+pBS sample in Jadhav21, and blue dots represent the BSS in Rain21}
    \label{fig: relation}
\end{figure}

We compared our results to those of Rain21 and Jadhav21. A sample of 138 new BSS within 50 open clusters were categorized; these stars have not been reported in previous works. 
The number of BSS in Galactic open clusters is increased by about 10\%, and the number of open clusters hosting BSS is increased by nearly 17\%.
As shown in Fig. \ref{fig: relation}, the age range of open clusters with blue stragglers in this work is consistent with that of Jadhav21; and the relationship between the number of blue stragglers and the cluster age is also similar to that of Jadahav21, the older open clusters show a trend of having more blue stragglers.

The properties of the 50 open clusters are listed in Table \ref{tab:paramters}, and the properties of BSS, YSS, and all other member stars are listed in Table \ref{tab:example_BSS}. 
The full tables and a complete list of cluster member stars are available at the CDS.
\subsection{Table 1: Open cluster properties}
\label{subsec:first_table}
We took advantage of the exquisite astrometric precision of Gaia DR3 to revisit the memberships of our selected clusters in a larger area around their center. Referring to the parameters of sample clusters such as the center positions, mean proper motions, and parallaxes provided in the literature \citep{2019ApJS..245...32L,2020A&A...635A..45C,2021MNRAS.502L..90F,2021A&A...646A.104H,2021RAA....21...93H,2022ApJS..260....8H}, we query Gaia Archive\footnote{\url{https://gea.esac.esa.int/archive/}} a cone of 20 pc radius (in most cases larger than the literature values) around the center of each OC. The membership estimation in the cluster region was then performed using pyUPMASK (for details of the process, see Sect.~\ref{sec:Membership}). Stars with a membership probability greater than 0.5 are regarded as member stars. 

For each cluster, we calculated its central coordinates, the proper motion, and the parallax parameters by taking the average values of the stars with $ \rm P_{m}>50\%$.
The age, distance module, reddening, and metallicity of clusters were derived from the best fitting isochrone (see step 2 in Sect. \ref{sec:search_bss}).
There are 31 of the 50 open clusters that have parameters from isochrone fitting in the literature. 
For these 31 open clusters, the parameters in this work are similar to those in the literature, except for a few OCs for which we obtained older ages.
The cluster names are mostly taken from Dias2021, otherwise the clusters' original names in the literature are presented.
On the basis of the morphology of the cluster CMDs, we divided the clusters into two categories and used the value of ``FLAG'' in Table~\ref{tab:paramters} to distinguish them. 
The clusters that have a clear main sequence pattern in the CMD are defined as Class 1 objects (FLAG = 1), and are shown in Figure~\ref{fig:CDM_class1} as an example. The clusters with a dispersed form of main sequence are ranked as Class 2 objects (FLAG = 2), as shown in Figure~\ref{fig:CDM_class2}.

\addtolength{\tabcolsep}{-2pt}
\begin{table*}[!th]
    \centering
    \caption{List of all clusters with newly identified BSS.\label{tab:paramters}}
    \begin{tabular}{rrrrrrrrrrrrrrr}
        \hline\hline
        idx & RA & DEC & pmRA & pmDE & Plx & $N_{mem}$ & $N_{BS}$ & $N_{pBS}$ & log(t) & DM & E(G-RP) & Z & name & FLAG \\
        \hline
        -- & deg & deg & mas/yr & mas/yr & mas &-- &-- &-- &--& mag& mag & dex &-- &--\\
        \hline
        1 & 29.40 & 37.76 & 9.69 & -11.97 & 2.28 & 467 & 1 & 0 & 9.30 & 8.1 & 0.0  & 0.015 & NGC\_752 & 1\\
        2 & 45.51 & 47.96 & 0.53 & -1.26 & 0.34 & 73 & 1 & 0 & 9.40 & 11.9 & 0.1  & 0.019  & CWNU\_365 & 1\\
        3 & 51.16 & 54.91 & 0.29 & -1.08 & 0.42 & 172 & 1 & 1 & 9.50 & 12.9 & 0.43 & 0.019  & CWNU\_483 & 2\\
        4 & 56.32 & 50.74 & 0.81 & -1.24 & 0.35 & 150 & 1 & 0 & 9.40 & 13.0 & 0.37 & 0.019  & CWNU\_17 & 2\\
        5 & 88.63 & 31.18 & 0.76 & -3.38 & 0.57 & 94 & 2 & 0 & 8.85 & 11.4 & 0.16 & 0.011  & CWNU\_197 & 1\\
        6 & 101.97 & -7.14 & -0.10 & 1.78 & 0.43 & 75 & 1 & 0 & 9.20 & 12.5 & 0.14  & 0.019  & CWNU\_251 & 1\\
        7 & 111.58 & -18.44 & -2.08 & 1.93 & 0.24 & 206 & 1 & 5 & 9.2 & 13.2 & 0.25  & 0.015 & FSR\_1253 & 2\\
        8 & 111.76 & -37.52 & -1.21 & 2.73 & 0.09 & 254 & 2 & 2 & 9.40 & 15.5 & 0.2  & 0.007 & DC\_3 & 2\\
        9 & 112.13 & -20.11 & -2.28 & 2.53 & 0.29 & 235 & 1 & 0 & 8.80 & 13.1 & 0.2  & 0.019 & LP\_198  & 1\\
        10 & 122.84 & -31.95 & -2.64 & 3.19 & 0.21 & 323 & 2 & 0 & 9.05 & 13.2 & 0.14  & 0.019  & LP\_386 & 1\\
        11 & 146.49 & -52.49 & -1.64 & -0.43 & 0.97 & 213 & 1 & 0 & 9.60 & 10.5 & 0.14  & 0.03  & LP\_2236 & 1\\
        12 & 153.20 & -60.90 & -6.37 & 3.11 & 0.50 & 281 & 1 & 0 & 8.60 & 11.5 & 0.07  & 0.017 & LP\_2059  & 1\\
        13 & 156.26 & -72.55 & -6.89 & 1.45 & 0.53 & 463 & 4 & 3 & 9.45 & 11.5 & 0.07  & 0.011  & UBC\_274 & 1\\
        14 & 170.24 & -59.30 & -5.25 & 1.81 & 0.35 & 302 & 2 & 0 & 9.20 & 12.5 & 0.11  & 0.019  & UBC\_271 & 1\\
        15 & 191.00 & -58.08 & -2.21 & -1.26 & 0.52 & 258 & 1 & 0 & 9.10 & 11.9 & 0.19 & 0.03  & UBC\_288 & 2\\
        16 & 233.85 & -57.67 & -3.61 & -5.75 & 0.40 & 177 & 1 & 0 & 9.90 & 12.2 & 0.43 & 0.019  & UFMG34 & 2\\
        17 & 236.04 & -55.26 & -6.06 & -4.39 & 0.36 & 167 & 5 & 2 & 10.0 & 13.2 & 0.72 & 0.015  & UFMG42 & 2\\
        18 & 236.57 & -56.81 & -1.68 & -3.23 & 0.38 & 378 & 1 & 2 & 9.10 & 12.1 & 0.33 & 0.019  & UBC\_306 & 2\\
        19 & 237.58 & -55.96 & -4.41 & -3.09 & 0.37 & 892 & 1 & 1 & 9.60 & 12.1 & 0.52 & 0.015  & UBC\_308 & 2\\
        20 & 241.49 & -52.68 & -1.70 & -2.69 & 0.30 & 453 & 2 & 1 & 9.60 & 12.5 & 0.57 & 0.019  & UFMG38 & 2\\
        21 & 244.94 & -49.97 & -3.13 & -2.47 & 0.39 & 181 & 1 & 0 & 9.55 & 11.7 & 0.35 & 0.011  & CWNU\_220 & 2\\
        22 & 245.71 & -50.18 & -2.38 & -4.12 & 0.39 & 332 & 1 & 0 & 9.20 & 12.5 & 0.37 & 0.03  & UBC\_543 & 2\\
        23 & 249.65 & -45.69 & 0.97 & -0.47 & 0.53 & 350 & 1 & 0 & 8.90 & 11.3 & 0.21 & 0.019  & CWNU\_244 & 1\\
        24 & 256.98 & -44.15 & -2.13 & -5.49 & 0.53 & 800 & 2 & 4 & 9.70 & 12.2 & 0.55 & 0.015  & UBC\_551 & 2\\
        25 & 259.26 & -41.69 & -0.04 & -3.58 & 0.61 & 228 & 1 & 0 & 9.10 & 11.8 & 0.44 & 0.009  & CWNU\_1 & 2\\
        26 & 261.58 & -37.79 & 0.54 & -0.99 & 0.45 & 211 & 1 & 0 & 9.30 & 11.4 & 0.4 & 0.011  & PHOC\_18 & 2\\
        27 & 264.87 & -35.08 & 0.28 & -3.68 & 0.31 & 109 & 3 & 3 & 9.80 & 12.8 & 0.4 & 0.015  & UFMG91 & 2\\
        28 & 264.91 & -33.23 & 0.14 & -0.76 & 0.36 & 496 & 1 & 0 & 9.60 & 11.5 & 0.55 & 0.019  & UBC\_567 & 2\\
        29 & 266.57 & -29.16 & 1.73 & -0.79 & 0.48 & 463 & 0 & 4 & 9.10 & 11.7 & 0.33 & 0.03  & UBC\_335 & 2\\
        30 & 267.68 & -30.21 & -0.11 & -1.33 & 0.33 & 788 & 0 & 3 & 9.00 & 11.9 & 0.39 & 0.015  & UBC\_570 & 2\\
        31 & 269.47 & -29.15 & -0.07 & -2.58 & 0.32 & 88 & 1 & 0 & 9.20 & 12.8 & 0.33 & 0.019  & UFMG93 & 1\\
        32 & 274.56 & -25.08 & -1.82 & 0.47 & 0.38 & 41 & 2 & 0 & 9.60 & 12.0 & 0.2 & 0.015  & UFMG46 & 1\\
        33 & 276.52 & -15.04 & -1.40 & -2.27 & 0.52 & 359 & 3 & 0 & 9.60 & 11.1 & 0.31 & 0.019  & CWNU\_132 & 2\\
        34 & 276.73 & -12.51 & 0.83 & 0.87 & 0.46 & 159 & 4 & 7 & 9.80 & 12.2 & 0.44 & 0.011  & CWNU\_37 & 2\\
        35 & 277.95 & -10.10 & 0.56 & -1.37 & 0.47 & 223 & 3 & 0 & 9.60 & 11.5 & 0.48 & 0.017  & CWNU\_70 & 2\\
        36 & 278.32 & -13.05 & -0.91 & -0.24 & 0.38 & 128 & 2 & 0 & 9.45 & 11.9 & 0.3 & 0.015  & CWNU\_397 & 1\\
        37 & 280.22 & -9.67 & 0.75 & -3.71 & 0.40 & 139 & 1 & 3 & 10.00 & 12.3 & 0.44 & 0.015 & CWNU\_404 & 2\\
        38 & 280.57 & -7.23 & 0.90 & -2.84 & 0.39 & 105 & 3 & 1 & 9.25 & 12.6 & 0.29 & 0.009  & PHOC\_6 & 1\\
        39 & 282.11 & 8.72 & -0.08 & -1.48 & 0.43 & 212 & 1 & 4 & 9.50 & 11.5 & 0.36 & 0.019  & UBC\_117 & 2\\
        40 & 282.34 & 3.35 & 1.75 & -1.28 & 0.60 & 185 & 3 & 1 & 9.90 & 12.2 & 0.75 & 0.019  & CWNU\_245 & 2\\
        41 & 286.16 & 13.42 & -0.26 & -4.54 & 0.52 & 81 & 3 & 3 & 9.50 & 11.4 & 0.32 & 0.019  & CWNU\_235 & 2\\
        42 & 289.42 & 8.69 & -2.96 & -5.27 & 0.61 & 148 & 2 & 0 & 9.55 & 12.0 & 0.35 & 0.015  & PHOC\_35 & 1\\
        43 & 296.40 & 21.16 & -0.34 & -3.57 & 0.46 & 259 & 2 & 2 & 9.70 & 11.9 & 0.41 & 0.019  & UBC\_127 & 2\\
        44 & 298.26 & 30.83 & -1.77 & -3.04 & 0.26 & 171 & 1 & 0 & 9.20 & 13.5 & 0.30 & 0.019  & CWNU\_319 & 1\\
        45 & 303.36 & 32.13 & -3.81 & -5.39 & 0.46 & 112 & 0 & 2 & 9.50 & 11.7 & 0.49 & 0.011  & HE\_40 & 1\\
        46 & 304.41 & 42.12 & -1.72 & -3.3 & 0.57 & 178 & 1 & 1 & 9.90 & 12.4 & 0.20 & 0.011  & CWNU\_106 & 2\\
        47& 305.80 & 41.69 & -3.64 & -8.35 & 0.82 & 247 & 2 & 0 & 9.10 & 10.6 & 0.30 & 0.011  & UBC\_142 & 2\\
        48 & 306.72 & 43.51 & -0.02 & -3.09 & 0.55 & 91 & 2 & 2 & 9.65 & 12.3 & 0.42 & 0.015 & CWNU\_439 & 2\\
        49 & 336.13 & 64.93 & -0.65 & -3.35 & 1.04 & 85 & 1 & 0 & 9.10 & 10.2 & 0.27 & 0.011 & UPK\_194  & 1\\
        50 & 352.50 & 62.34 & -3.15 & -2.98 & 0.60 & 134 & 1 & 0 & 9.85 & 11.7 & 0.40 & 0.017 & CWNU\_109 & 2\\
       \hline 
    \end{tabular}
\tablefoot{The listed parameters for each column are as follows: sequence number (idx), the equatorial coordinates (RA, DEC at J2016.0), mean proper motions (pmRA, pmDE), mean parallaxes (Plx), number of member stars ($N_{mem}$), number of blue straggler stars ($N_{BS}$), number of possible blue straggler stars ($N_{pBS}$), the logarithm of the age (log(t)), distance module (DM), reddening (E(G-RP)), metallicity (Z), cluster names (name), and the ``FLAG.''}
\end{table*}

\subsection{Table 2: Stragglers in open clusters}
\label{subsec:second_table}

Table~\ref{tab:example_BSS} describes the catalog of BS, pBS, YSS, and all the other member stars.
A complete list of these stars is available online.

\addtolength{\tabcolsep}{-3pt}
\begin{table*}[!th]
    \centering
    \caption{Description of the catalog of BSS, pBSSs, and all the other member stars.}
    \label{tab:example_BSS}
    \resizebox{\linewidth}{!}{
    \begin{tabular}{llll}
    \hline \hline
        Column & Format & Unit & Description \\
       \hline
        ID\_DR3 & Long & -- & Unique source designate in Gaia DR3\\
        cluster & String & -- & The name of the corresponding cluster \\
        $\rm P_{m}$ & Float & -- & The membership probability of stars\\
        f\_strg & String & -- & Flag indicating that the star is a genuine blue straggler (BS), a possible blue straggler candidate (pBS), a yellow straggler (YSS), or not a straggler (N) \\
        RA & Float & deg & Right ascension at Ep=2016.0\\
        DEC & Float & deg & Declination at Ep=2016.0\\
        $ \rm \mu_{\alpha^\ast}$ & Float & mas  $\rm yr^{-1}$ & Proper motion in right ascension direction\\
        $ \rm \mu_{\alpha^\ast}\_err$ & Float & mas  $\rm yr^{-1}$ & Standard error of $ \rm \mu_{\alpha^\ast}$\\
        $ \rm \mu_{\delta}$ & Float & mas  $\rm yr^{-1}$ & Proper motion in declination direction\\
        $ \rm \mu_{\delta}\_err$ & Float & mas  $\rm yr^{-1}$ & Standard error of $ \rm \mu_{\delta}$\\
        $ \rm \varpi$ & Float & mas & Parallax \\
        $ \rm \varpi \_err$ & Float & mas & Standard error of parallax \\
        G\_mag & Float & mag & G band mean magnitude\\
        BP\_mag & Float & mag & Integrated BP band mean magnitude\\
        RP\_mag& Float & mag & Integrated RP band mean magnitude\\
        Gmag\_cor & Float & mag & G band magnitude after differential reddening correction\\
        (G-RP)\_cor & Float & mag & (G-RP) magnitude after differential reddening correction \\
       \hline 
    \end{tabular}
    }
\tablefoot{This table contains 17 columns, including the astrometric and photometric parameters of cluster members in Gaia DR3 (Col. 1, Cols. 5--15), the corresponding cluster name (Col. 2), the membership probability of these stars in this work (Col. 3), the results of identification of stragglers (Col. 4), and the magnitudes after differential reddening correction (Cols. 16--17).} 
\end{table*}

\section{Discussion}
\label{sec:discus}

\subsection{DC 3: Globular cluster candidate with BSS}
\label{subsec:DC_3}

DC 3 was first identified as an open cluster by \cite{2005A&A...435..545D}, and later listed in the major open cluster catalogs (e.g., \citealt{2013A&A...558A..53K} and \citealt{2021MNRAS.504..356D}).
As described in Sect. \ref{sec:Membership}, we have homogeneously determined cluster members by pyUPMASK using Gaia DR3 data.
For DC 3, we obtained 254 member stars with $ \rm P_{m}> 0.5,$ and identified two BSS and three pBSS that were not reported in previous studies. \cite{2021MNRAS.504..356D} also provided the membership estimation for DC 3 using Gaia DR2 data.

\begin{figure}[!htbp]
    \centering
    \includegraphics[width=\linewidth]{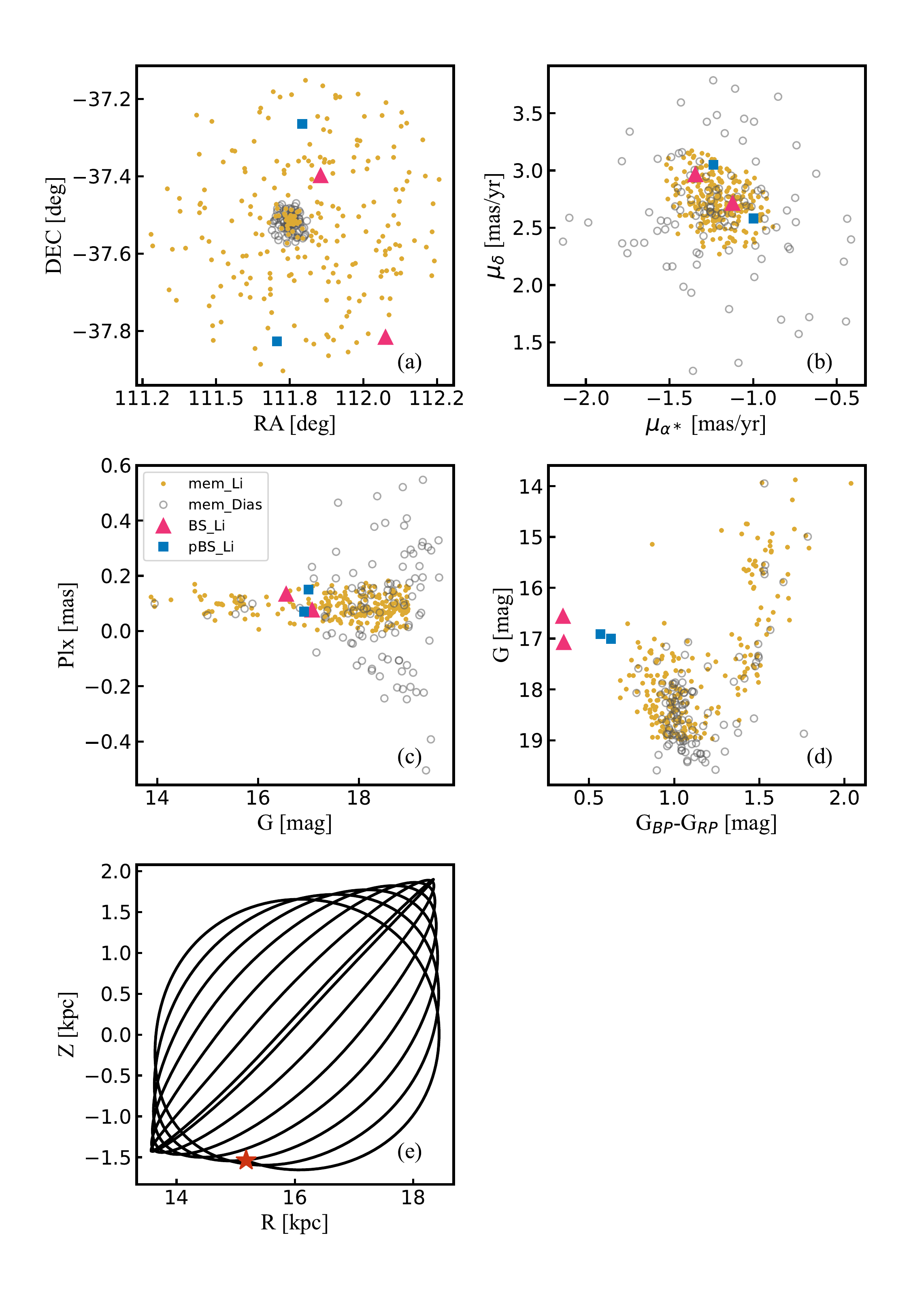}
    \caption{Comparison between our result for DC 3 members and that from \cite{2021MNRAS.504..356D}, and the orbit of DC 3 in galactocentric coordinates calculated in this work. In panels (a) to (d) the yellow dots represent the member stars of DC 3 with $ \rm P_{m}>0.5$ in this work, the magenta triangles and blue squares are respectively the BSs and pBSs identified in this work, and the gray circles are the cluster members from \cite{2021MNRAS.504..356D} with member probabilities over 0.5. Panel (a): Spatial distribution for the cluster members and identified blue straggler candidates. (b): Proper motion distribution of cluster members and identified blue straggler candidates. (c): Parallax distribution for cluster members and identified blue straggler candidates. (d): CMD of DC 3. (e): Orbit of DC 3 in Galactocentric coordinates projected in the R-Z plane, where $\rm R=\sqrt{X^2+Y^2}$. The red star represents the present-day position.}
    \label{fig:DC3_5diag}
\end{figure}

Panels (a) to (d) of Fig. \ref{fig:DC3_5diag} show the comparison between our estimated cluster members and that from \cite{2021MNRAS.504..356D}.
Compared with the \cite{2021MNRAS.504..356D} sample, our cluster members are distributed in a much larger area, while appearing to be more compact in the parallax distribution and proper motion space as well.
The BS candidates we identified are found outside the confined cluster region inspected by \cite{2021MNRAS.504..356D}.

Thanks to the much more radial velocities provided by Gaia DR3 \cite{2022arXiv220605902K}, we can further analyze the kinematic properties of DC 3.
Table~\ref{tab:DC 3 params} lists the cluster parameters from \cite{2013A&A...558A..53K} (hereafter K13), \cite{2021MNRAS.504..356D} (hereafter D21), and this work (hereafter L23).
Compared with K13 and D21, we obtained the radial velocity data of 15 member stars for the first time, and the metallicity of the cluster we fitted is more metal poor than that of D21.

We calculated the orbit of the cluster under the gravitational potential of the Milky Way \citep{2018PASP..130k4501M} using the galpy\footnote{A Python library for galactic dynamics \url{http://github.com/jobovy/galpy}} package \citep{2015ApJS..216...29B} based on the position and velocity phase space data of DC 3.
The panel (e) of Fig. \ref{fig:DC3_5diag} shows the trajectory of DC 3 in the galactocentric cylindrical coordinates (R vs. Z).
The time duration of the trajectory is 2.5 Gyr, corresponding to the age derived from isochrone fitting, and the red star is the current position of DC 3.
As we can see from the panel, the maximum distance of DC 3 to the Galactic plane can reach about 1.8 kpc, and the galactocentric distance can reach as far as 18 kpc.

In this work, DC 3 shows quite different characteristics from most of the open clusters in the Milky Way, including the older age, metal-poor properties, the morphology of CMD, and its large distance to the Galactic plane and faraway apocenter on the Galactic disk. 
As a consequence, we propose that DC 3 is more suitable to be classified as a sparse galactic globular cluster.

\begin{table}[!th]
    \caption{Parameters of DC 3.}
    \label{tab:DC 3 params}
    \centering
    \resizebox{\linewidth}{!}{
    \begin{tabular}{rccccc}
        \hline\hline
         Param & Unit & $ \rm Value_{K13}$ & $ \rm Value_{D21}$ & $ \rm Value_{L23}$ & $ \rm Num_{L23}$ \\
        \hline
        RV & km $\rm s^{-1}$ & -- & -- & $103.89\pm 18.60$ & 15 \\
        $\rm V_{t}$ & km $\rm s^{-1}$ & 107.81 &109.61 & $121.17\pm 8.07 $ & 254 \\
        $\rm [Fe/H]$ & dex &-- & $- 0.146\pm 0.049$ & $-0.455$ & --\\
        Dist & kpc & 4.32 & 7.93 & 8.59  & -- \\
        logt & -- & 9.26 & 9.47 & 9.40 & -- \\
        \hline
        \end{tabular}
        }
    \tablefoot{Columns 3--5 are the values of the parameters in Col. 1 from different works (see text). RV and $ \rm V_{t}$ are the radials and tangential velocities of the cluster respectively. $ \rm [Fe/H]$, Dist, and logt are the isochrone fitting results for metallicity, photometric distance, and age of DC 3, respectively. In this work, RV and $ \rm V_{t}$ are calculated by the average radial and tangential velocity of the cluster members with $ \rm P_{m}>0.5$. The number of members included in the calculation of these two parameters is listed in column 6.}
\end{table}

\subsection{Five supplementary OCs from CG20}
\label{subsec:CG20_add}

For the open clusters in CG20a or CG20b that had not been picked out by Rain21 or Jadhav21 as clusters with blue straggler candidates, we also re-determined cluster memberships following the method described in Sect. \ref{sec:Membership}.
After the steps described in Sect. \ref{sec:search_bss}, we recognized five open clusters with blue stragglers that had not been listed in the catalog of Rain21 or Jadhav21: NGC 752, FSR 1253, LP 198, LP 2059, and UPK 194. 
The CMDs of these clusters are shown in the Appendix.

\begin{figure}[!htbp]
    \centering
    \includegraphics[width=\linewidth]{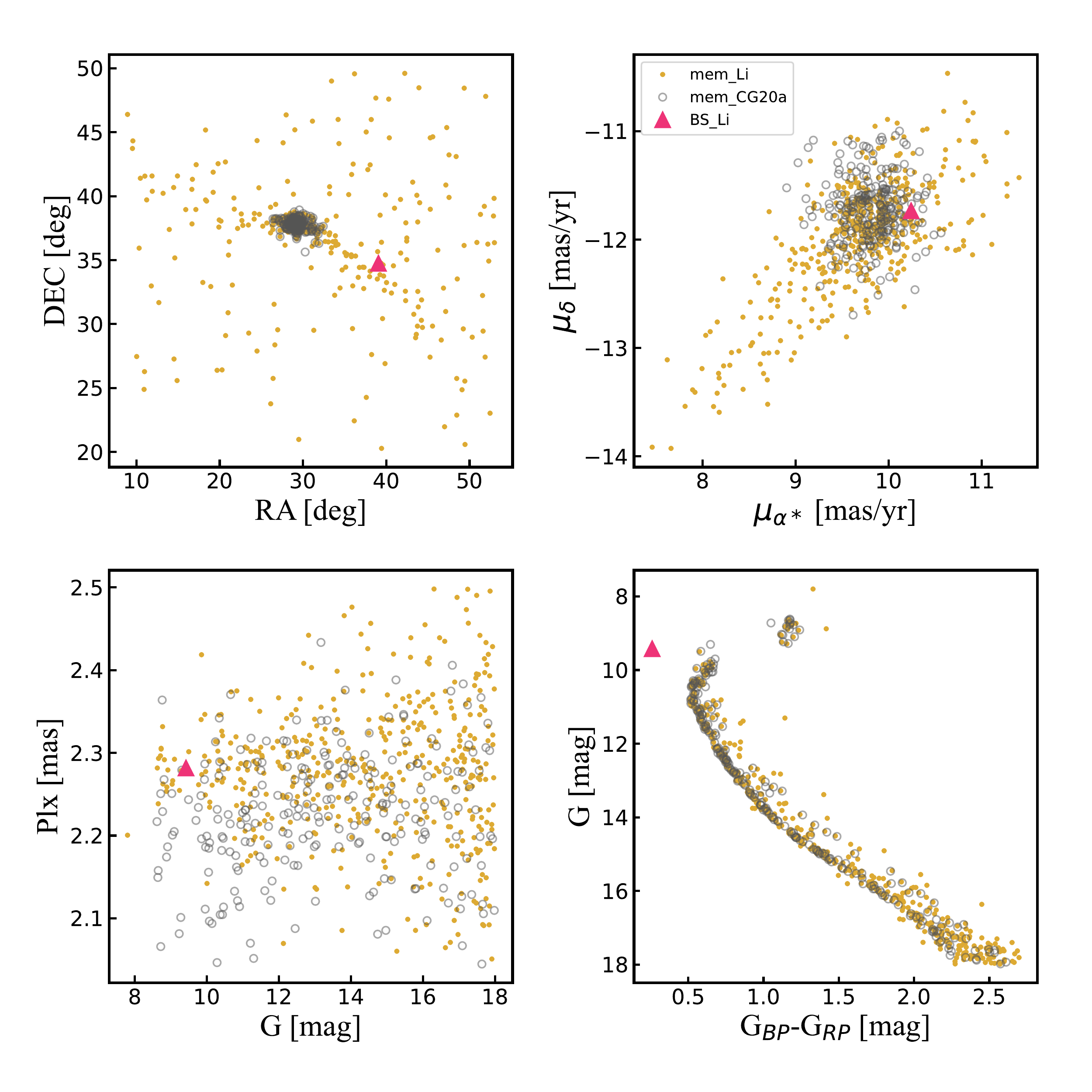}
    \caption{Parameters of NGC 752. Yellow dots are the cluster members re-determined in this work, and a magenta triangle represents the BS of NGC 752 we identified. The member stars provided by \cite{Cantat-Gaudin_2020} are shown as gray circles. All these stars have a member probability over $50\%$.}
    \label{fig: NGC752}
\end{figure}

It is worth noting that for the first time we have found a blue straggler on the tidal tail of NGC 752.
NGC 752 is a close open cluster at a distance of about 440 pc \citep{2013A&A...558A..53K, 2020A&A...635A..45C, 2021MNRAS.504..356D}. 
Its radius in \cite{2013A&A...558A..53K} is $ \rm r2 = 1.4$ deg, while in \citet{Cantat-Gaudin_2020} and \citet{2021MNRAS.504..356D} its half-number radius is about 0.47 deg. 
In a recent work \citet{2021MNRAS.505.1607B} found tidal tails that are twice as long on both sides of NGC 752, as previously reported, by using Gaia EDR3.
The area covered by NGC 752 members extended to 260pc in diameter, corresponding to an angular scale of about 37 deg. 
Inspired by this work, we expanded our investigated area of NGC 752 to an angular radius of about 19 deg, or a radius of 150 pc at the distance of NGC 752. 
We performed the pyUPMASK algorithm and the specific features of resulting members with probability $ \rm P_{m}>0.5$ are shown in Fig. \ref{fig: NGC752}, together with those data provided by \cite{Cantat-Gaudin_2020}. 
As shown in the figure, there is a blue straggler on the tidal tail of NGC 752, which has never been reported. 
A detailed study of this blue straggler will be presented in our next work (Li et al. in preparation).

\section{Summary}
\label{sec:summary}

We searched for blue stragglers from over 1000 open clusters, either newly found open clusters based on Gaia data or those beyond the sample clusters picked out by Rain21 and Jadhav21.

First, we implemented a uniform membership determination for cluster samples in a larger sky coverage to perform a more comprehensive search for blue straggler members in open clusters.
In order to identify blue straggler stars more reliably and accurately, we used five dimension parameters (position, proper motion, and parallax) provided by Gaia DR3 to calculate the membership probabilities of stars with pyUPMASK, and selected stars with $ \rm P_{m}>0.5$ for the subsequent analysis.

Then we re-evaluated the properties of the clusters by isochrone fitting, and picked out the blue stragglers in a specific region of the clusters' CMD.
In view of the different reliabilities, the blue stragglers we identified were divided into two different classes (BS and pBS) based on their relative positions to MSTO.
As a result, we identified 138 blue straggler candidates from 50 open clusters, including 81 BSS and 57 pBSSs, which had not been reported before. In addition, 14 YSS in seven clusters of our sample were also recognized.

From our results, we demonstrated that DC 3, previously assigned as an open cluster, should in fact be a globular cluster with BSS, in accordance with its older age, mental-poor properties, the morphology of CMD, and orbital property.
In this work we also recognized eight more blue stragglers in five open clusters from CG20a or CG20b that are not listed in the catalogs of Rain21 and Jadhav21.

\begin{acknowledgements}
 
We thank the anonymous referee for the instructive comments and suggestions which help us a lot to improve the paper. Li Chen acknowledges the support from the National Natural Science Foundation of China (NSFC) through the grants 12090040 and 12090042. 
Jing Zhong would like to acknowledge the NSFC under grants 12073060, and the Youth Innovation Promotion Association CAS.
We acknowledge the science research grants from the China Manned Space Project with NO. CMS-CSST-2021-A08.

This work has made use of data from the European Space Agency (ESA) mission Gaia (\url{https://www.cosmos.esa.int/gaia}), processed by the Gaia Data Processing and Analysis Consortium (DPAC,\url{https://www.cosmos.esa.int/web/gaia/dpac/consortium}). Funding for the DPAC has been provided by national institutions, in particular the institutions participating in the Gaia Multilateral Agreement.
\end{acknowledgements}

\bibliographystyle{aa}
\bibliography{bib}

\begin{appendix} 
\section{The CMDs of five supplementary OCs}

\begin{figure}[!htbp]
    \centering
    \includegraphics[width=\linewidth]{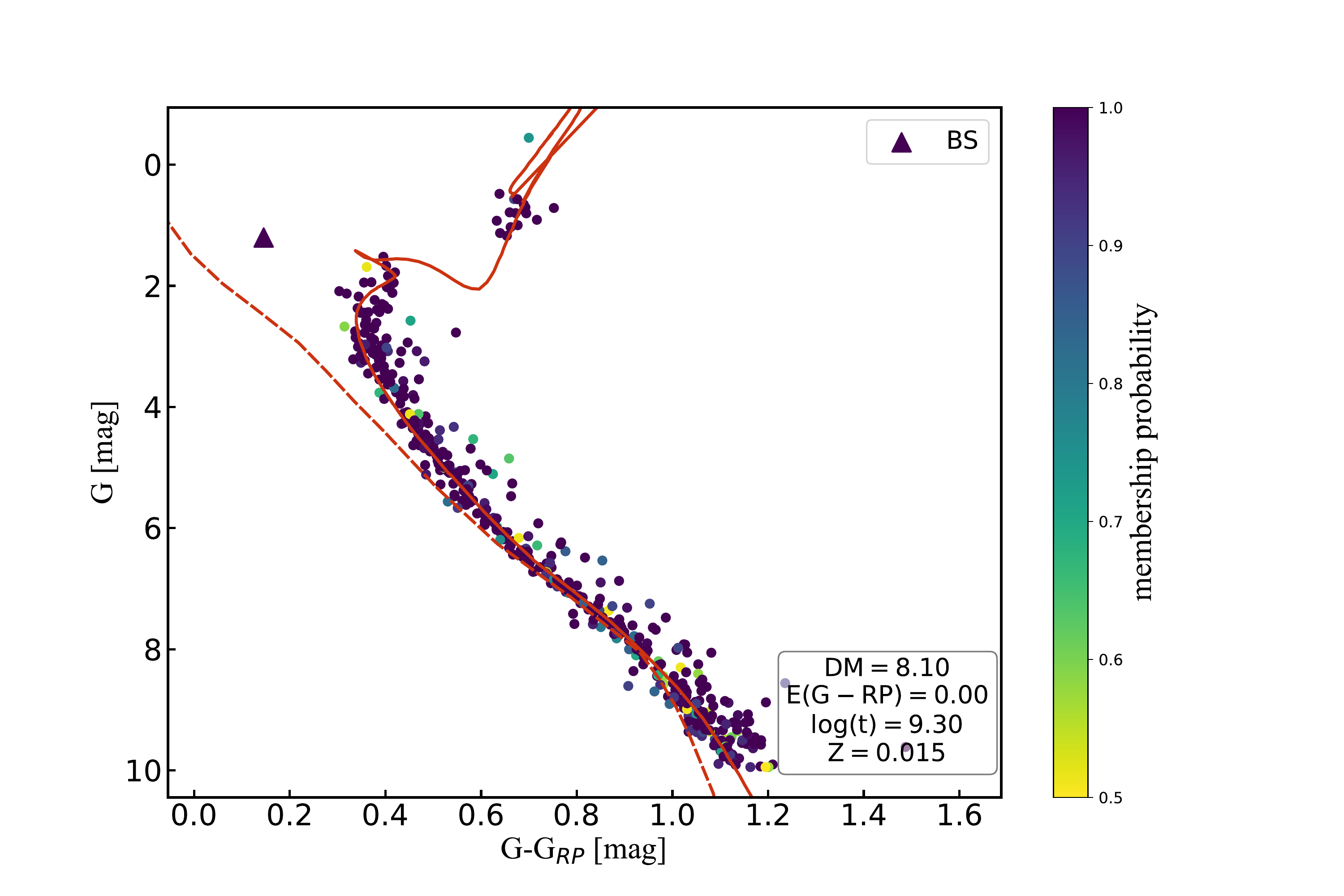}
    \caption{Same as Fig. \ref{fig:CDM_class1}, but for cluster NGC 752.}
\end{figure}

\begin{figure}[!htbp]
    \centering
    \includegraphics[width=\linewidth]{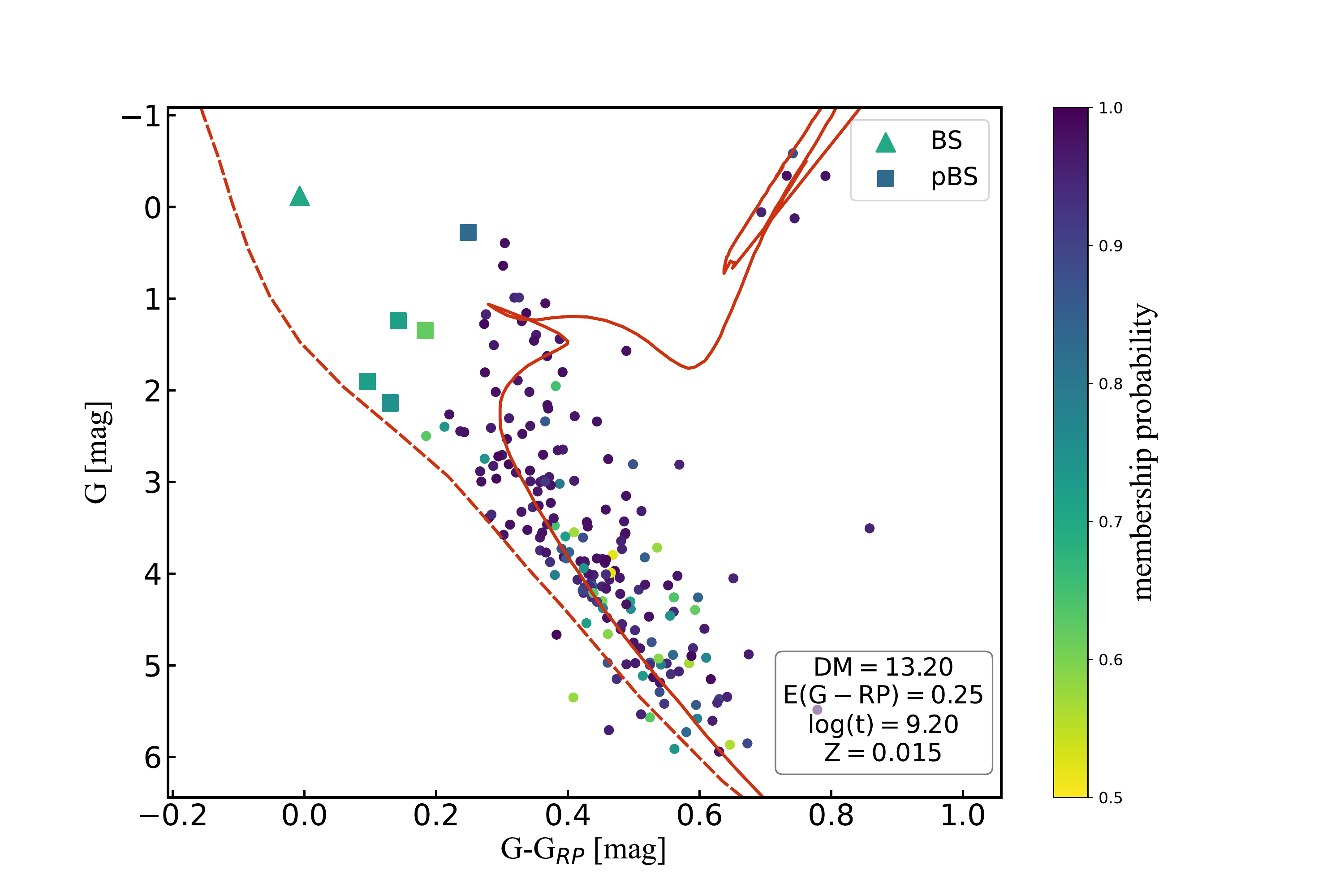}
    \caption{Same as Fig. \ref{fig:CDM_class1}, but for cluster FSR 1253.}
\end{figure}

\begin{figure}[!htbp]
    \centering
    \includegraphics[width=\linewidth]{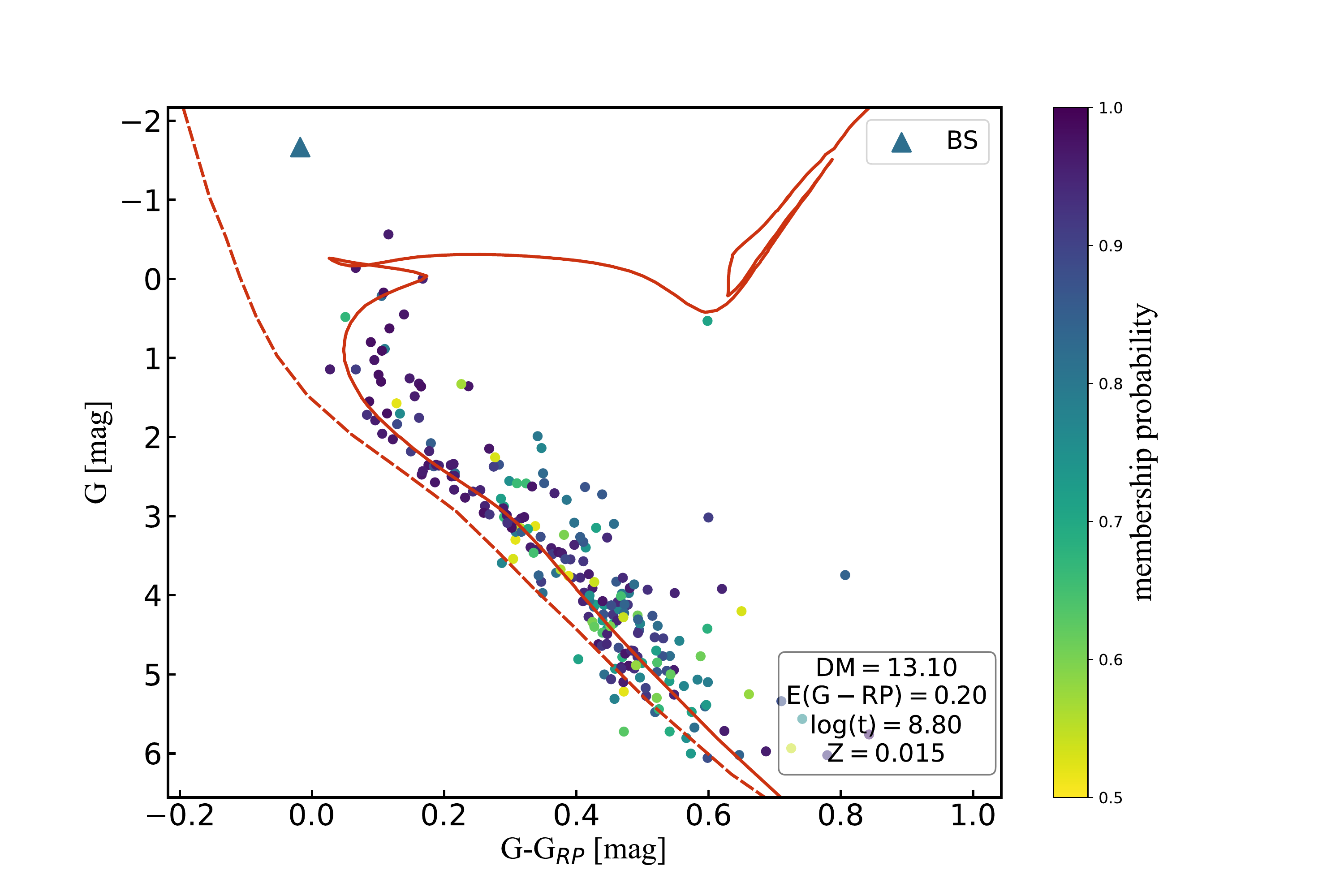}
    \caption{Same as Fig. \ref{fig:CDM_class1}, but for cluster LP 198.}
\end{figure}

\begin{figure}[!htbp]
    \centering
    \includegraphics[width=\linewidth]{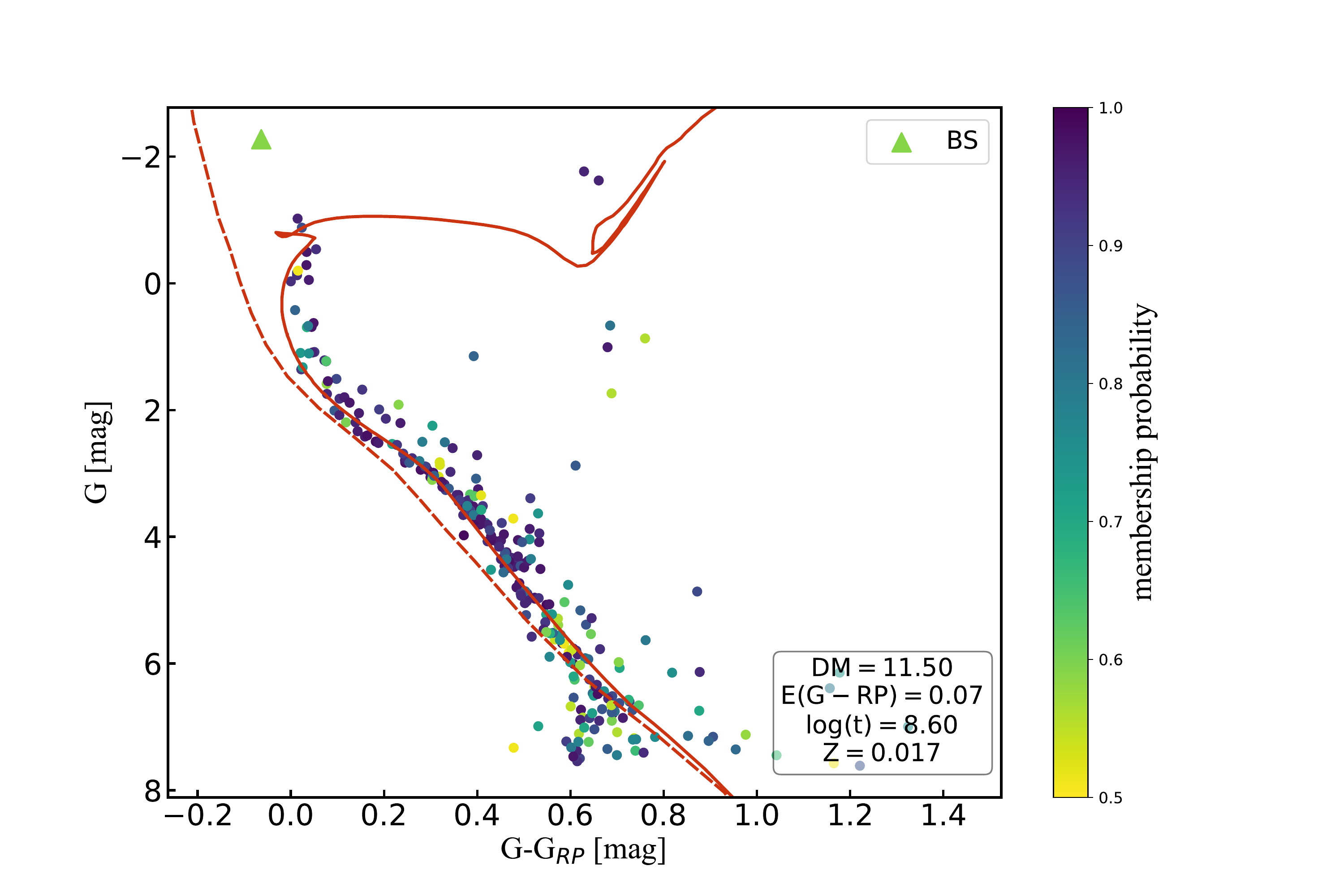}
    \caption{Same as Fig. \ref{fig:CDM_class1}, but for cluster LP 2059.}
\end{figure}

\begin{figure}[!htbp]
    \centering
    \includegraphics[width=\linewidth]{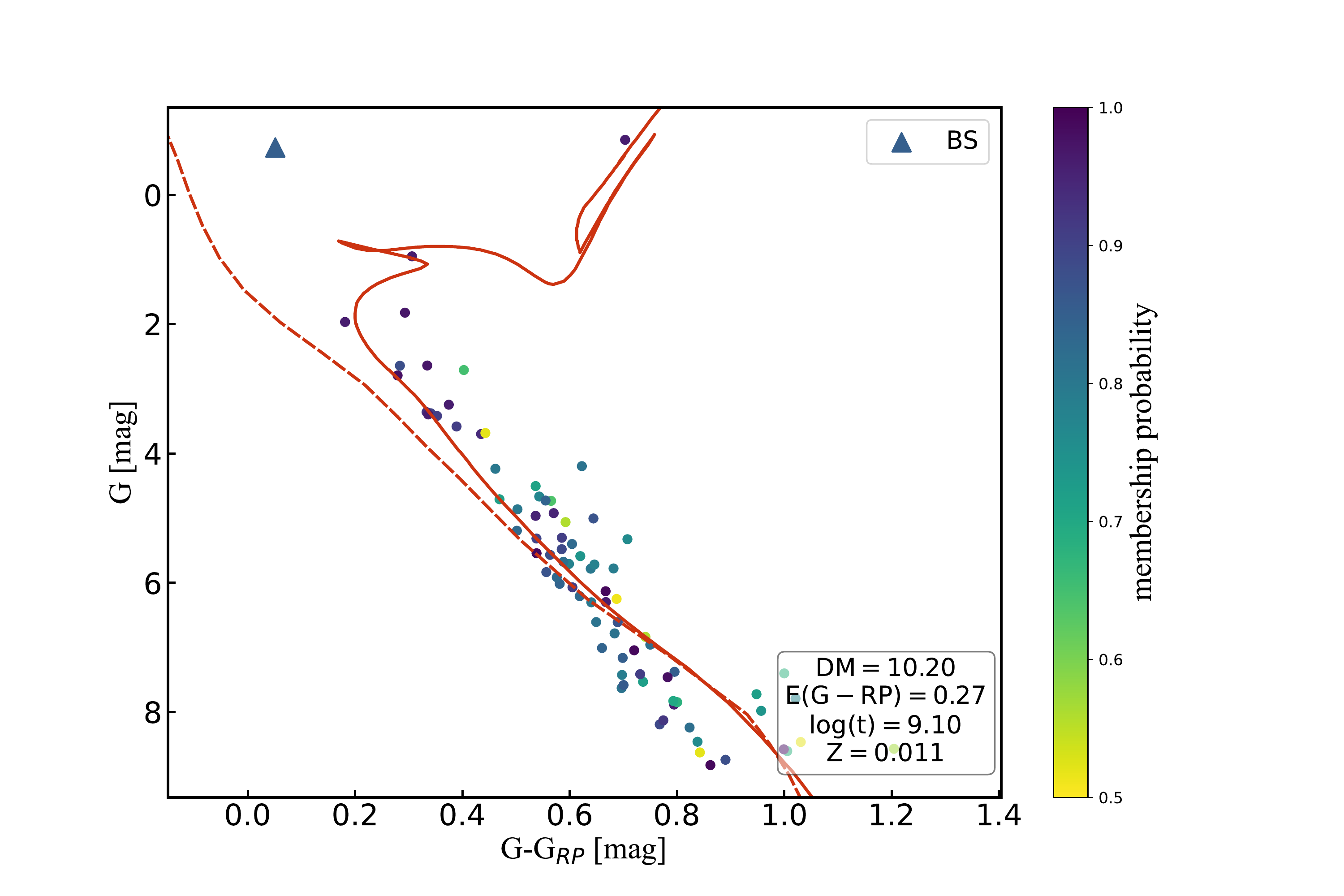}
    \caption{Same as Fig. \ref{fig:CDM_class1}, but for cluster UPK 194.}
\end{figure}

\end{appendix}

\end{CJK}
\end{document}